\documentclass[]{iopart}
\usepackage{iopams}  
\usepackage{graphicx, epsfig}
\begin{document}

\def\be{\begin{equation}}
\def\ee{\end{equation}}
\def\bea{\begin{eqnarray}}
\def\eea{\end{eqnarray}}
\def\SIC{{\it a}-SiC }

\title[]{Atomistic modeling of amorphous silicon carbide: An approximate first-principles study 
in constrained solution space} 

\author{Raymond Atta-Fynn}
\address{Department of Physics and Astronomy, The University of Texas, Arlington, TX 76019, USA} 
\ead{attafynn@uta.edu}

\author{Parthapratim Biswas}
\address{Department of Physics and Astronomy, The University of Southern Mississippi, 
Hattiesburg, MS 39406, USA} 
\ead{partha.biswas@usm.edu} 

\begin{abstract} 
Localized basis {\it ab initio} molecular dynamics simulation within the density functional framework 
has been used to generate realistic configurations of amorphous silicon carbide ({\it a}-SiC).  Our 
approach consists of constructing a set of smart initial configurations that conform essential geometrical 
and structural aspects of the materials obtained from experimental data, which is subsequently driven 
via first-principles force-field to obtain the best solution in a reduced solution space. A combination 
of {\it a priori} information (primarily structural and topological)  along with the ab-initio optimization 
of the total energy makes it possible to model large system size (1000 atoms) without compromising the 
quantum mechanical accuracy of the force-field to describe the complex bonding chemistry 
of Si and C. The structural, electronic and the vibrational properties of the models have been studied 
and compared to existing theoretical models and available data from experiments. We demonstrate that 
the approach is capable of producing large, realistic configurations of \SIC from first-principles simulation 
that display excellent structural and electronic properties of {\it a}-SiC. Our study reveals the presence of 
predominant short-range order in the material originating from heteronuclear Si--C bonds with coordination 
defect concentration as small as 5\% and the chemical disorder parameter of about 8\%. 
\end{abstract}

\pacs{81.05.Gc, 02.70.Ns, 71.15.Mb}

\submitto{\JPCM}


\section{Introduction}
Silicon carbide is a wide bandgap semiconductor having a range of applications from optoelectronic 
devices, gas turbines, heat exchangers to ceramic fans.  
The amorphous form of SiC is particularly interesting due to temperature stability of 
semiconducting properties making it possible to use under extreme condition in applications 
such as high-temperature engines, turbines and reactors~\cite{Levin, Emin, Marshall,
Connor, Pechenik, Thorpe, Priya}. There have been a number of studies of amorphous SiC in recent 
years emphasizing mostly on the structural aspect of the material. Of particular importance is 
the nature of the short range order (SRO) in the material. Despite numerous experimental and 
theoretical studies, the exact nature of the SRO and the extent of its presence are still missing.
While bonding in its various crystalline counterpart is $sp^{3}$ corresponding to heteronuclear 
Si--C bonds, the nature of bonding chemistry in \SIC is far more complex and is not well 
understood.  Different experimental probes on \SIC often appears to 
suggest contradictory results: from presence of purely heteronuclear Si--C bonding to an admixture of 
heteronuclear Si--C (chemical order) and homonuclear Si--Si and C--C bonds 
(chemical disorder). Unlike silicon that conforms primarily to tetrahedral $sp^{3}$ bonding, the 
complex carbon chemistry permits the atom to hybridize via $sp$, $sp^{2}$ and $sp^{3}$ bond 
formation in the amorphous environment. The structure of the materials also depends on the 
growth conditions and the subsequent treatment of the sample. As a result, various experimental 
probes often provide a very different microscopic picture of the material. 

In this communication we study the structural, electronic and vibrational properties of a 1000-atom 
model of \SIC via {\it ab initio} molecular dynamic simulation using a localized basis sets and pseudopotential 
density functional approach. Our emphasis is on developing an effective approach that models a 
sufficiently large system size without compromising the accuracy of the interaction among the atoms by focusing 
on the relevant part of the configuration space by using a suitable prior from experimental data. This 
has been achieved by generating a set of smart generic configurations and optimizing the resulting 
structural models in the reduced solution space of the first-principles force field. The paper is 
organized as follows.  In Section 2, we provide a brief discussion of earlier works and the present state of 
knowledge of the material in amorphous state.  Section 3 describes the essential idea behind our method and 
how to construct smart configurations using experimental information and some geometrical, topological and 
structural characteristics of the material.  The simulation procedure is outlined in Section 4, which is then 
followed by a discussion of the results in Section 5.

\section{Summary of earlier works} 
There has been a wealth of experimental information available from a range of studies 
including from Raman scattering, Electron and X-ray diffraction to Auger spectroscopy; 
however, none of these studies can provide a conclusive picture of the structural 
properties in the amorphous state. 
A number of experimental studies (such as Raman scattering~\cite{Gorman} and X-ray diffraction~\cite{Ishimaru1}) have 
indicated significant presence of chemical disorder via homonuclear bond formation, whereas there 
are evidence from Auger~\cite{Lee}, X-ray photoemission~\cite{katayama}, and extended X-ray absorption~\cite{Kaloyeros} 
experiments indicating presence of heteronuclear bonds.  
Kaloyeros \etal~\cite{Kaloyeros} performed extended X-ray absorption spectra and 
energy loss fine structure experiments to study sputtered \SIC films and 
concluded that the short range chemical order was essentially consisting 
of only Si--C heteronuclear bonds. 
An X-ray adsorption and Raman spectroscopy study of SiC amorphized by ion bombardment by 
Bolse~\cite{Bolse} concluded that the chemical short range order not only comprised of 
Si--C bonds but also homonuclear C--C and Si--Si bonds. Electron diffraction studies of SiC amorphized 
by ion beams confirms this observation~\cite{Ishimaru1}, and shows the recovery of the SRO upon annealing 
to 1073 K~\cite{Ishimaru2}. Fast neutron irradiation induced amorphization of SiC experiments by 
Snead \etal revealed  that the atomic density of \SIC 2.85 g/cm$^3$ at 300 K increased 
to 3 g/cm$^3$ after annealing at 1100 K due to the recovery of SRO~\cite{Snead}. The picture that 
emerges from these experimental results points to the fact that \SIC consists of both homo and heteronuclear 
bonds but it is difficult to quantify to what extent the short range chemical order is present in the 
network. 

As far as theoretical studies are concerned, a fair amount of work can be found 
in literature~\cite{Finnocchi, Kelires, Tersoff 89, Tersoff 94, Mura, Ivashchenko, Gao, 
Malerba, Yuan, Gao2, Brenner, Rino}. Finocchi and co-workers~\cite{Finnocchi} were the 
first to perform a quench-from-the-melt {\it ab initio} molecular dynamics study using a 54- and 
64-atom supercells and observed that 45\% of the total bonds formed by C were homonuclear, of which 
15\% of the bonds were 3-fold coordinated ($sp^{2}$) with bond angles close to 120$^\circ$ 
(making these regions essentially graphite-like) and that \SIC had negligible chemical 
ordering. A subsequent study by Kelires~\cite{Kelires} 
via quench-from-the-melt Monte Carlo simulations of a 216-atom \SIC sample using the empirical 
Tersoff potential~\cite{Tersoff 89} observed strong chemical ordering, but 
50\% of the C atoms were 3-fold coordinated and 33\% of the C bonds were homonuclear. 
In an attempt to resolve the discrepancy, Tersoff~\cite{Tersoff 94} performed classical MD simulations using an 
improved version of the empirical potential~\cite{Tersoff 89} introducing a measure of chemical disorder via the 
ratio
\be
\chi = \frac{n_{CC}}{n_{SiC}}
\label{eqn1}
\ee
where $n_{\mathrm CC}$ is the number of C--C bonds and $n_{\mathrm SiC}$ is the number of Si--C bonds. $\chi$=0 
implies no chemical disorder and $\chi$=1 indicates complete disorder in the network.  Based on 
this definition, the author concluded that if the graphite-like regions in Ref.~\cite{Finnocchi} 
were excluded then the combined results in Refs.~\cite{Finnocchi},~\cite{Kelires}~and~\cite{Tersoff 94} 
corresponded to $0.5\le\chi\le0.6$. Mura and co-workers~\cite{Mura} performed classical MD simulations 
for {\it a}-Si$_{x}$C$_{1-x}$ alloys in the composition range $0.125<x<0.875$ using the modified Tersoff 
potential and a 512-atom supercell. For {\it a}-Si$_{0.5}$C$_{0.5}$ they obtained $\chi=0.6$ and 
concluded that for high C contents, the value of $\chi$ is linked to a sizable distortion of the Si 
sublattice.  Ivashchenko \etal~\cite{Ivashchenko} performed 
$sp^{3}s^*$ tight-binding MD simulations for models of \SIC 
containing 54 to 216 atoms. They observed the existences of 
homonuclear bonds and coordination defect concentration of 6\% to 13\%.  
MD simulations of ion or neutron bombardment of \SIC have 
been performed by Gao and Weber~\cite{Gao} and by 
Malerba and Perlado~\cite{Malerba}. Gao and Weber~\cite{Gao} used the Tersoff 
potential~\cite{Tersoff 94} and obtained $\chi$=0.41 in the damaged region created by 
10 KeV Au recoils in a 250,000-atom supercell of 3C-SiC. Similarly, 
using classical MD simulations and a different version of the Tersoff 
potential~\cite{Tersoff 89}, Malerba and Perlado~\cite{Malerba} studied the 
damage accumulation process for 100 eV recoils 
in a 512-atom supercell of 3C-SiC embedded in a larger bounding crystal containing 
1216 atoms. They obtained $\chi=0.34$ for both the irradiated \SIC and the 
reference quench-from-the-melt model of {\it a}-SiC. However, $\chi$ 
reduced to 0.22 in the irrdiated model and remained at 0.34 for the 
reference quench-from-the-melt model. Since previous simulations 
using the Tersoff potential~\cite{Kelires,Tersoff 94} produced $\chi$=0.5--0.6, 
the rather different results ($\chi$=0.22--0.34) by Malerba and Perlado~\cite{Malerba} obtained 
using the same potential as described above is quite puzzling. However, 
Yuan and Hobbs have shown that if the amorphous-crystalline boundary is excluded, 
then the results of Malerba and Perlado~\cite{Malerba} will correspond to 
($\chi$=0.35--0.5). Quite recently, Devanathan \etal~\cite{Gao2} 
have performed classical MD simulations using a Brenner-type 
interatomic potential~\cite{Brenner}   
to model 4096-atom configuration of {\it a}-SiC. They observed that 
the density of the quenched liquid was in excellent agreement with 
experiment and $\chi$=0.06--0.13 for annealing temperatures ranging from 300 K to 2500 K.  
Finally, Rino \etal~\cite{Rino}, using an empirical potential with Coulomb, 
charge-dipole and Van der Waals interactions included in the two-body term, 
performed classical MD simulation for a large supercell of \SIC containing 10,648 
atoms and obtained $\chi$=0, in perfect agreement with the 
experiments by Kaloyeros~\etal ~\cite{Kaloyeros}. 

In summary, the bulk of the experimental and theoretical studies 
appear to favor a non-zero value of $\chi$ with the exception of the experimental work by 
Kaloyeros \etal ~\cite{Kaloyeros} and the theoretical models proposed by Rino \etal~\cite{Rino}. 
As mentioned before, experimental results indicate that bonding in amorphous SiC is 
a mixture of both homonuclear and heteronuclear bonding consisting of $sp^2$ and $sp^3$ 
hybridizations. Given the complex nature of bonding chemistry in the amorphous environment, it is not surprising 
that models based on empirical or semi-empirical potentials are at best incomplete to describe 
some of the structural and electronic properties of {\it a}-SiC. Since most of the theoretical models 
proposed so far are based on either empirical or semi-empirical potentials, it is important 
to carry a full scale first-principles study of \SIC so that the details of the quantum mechanical 
nature of the bonding between C and Si and the local chemistry can be taken into account in the 
model construction. While the models proposed by Finocchi \etal~\cite{Finnocchi} did take into account the 
quantum mechanical interactions within the density functional theory via local density approximation, 
the size of the simulation cell (typically 64 atoms) was too small to represent a realistic properties 
of bulk {\it a}-SiC. Since structural properties are directly linked with the electronic and vibrational properties of 
the materials, a predictive structural modeling should be done in conjunction with electronic and 
vibrational properties calculations for a reasonable system size, and to be compared with experiments. 
Finally, there are very few models in the literature that address the electronic properties 
of {\it a}-SiC~\cite{Finnocchi, Ivashchenko}. The present study attempts to provide a comprehensive 
picture of the structural and electronic properties of the alloy by selectively sampling the solution 
using the prior information obtained from experiments.

\section{Strategy of our approximate simulation: sampling solution space with prior information }

In conventional electronic structure problem one begins with a set of atomic coordinates and an interacting potential or Hamiltonian. The electronic density of states 
is obtained either by solving the Schr\"odinger equation within the first-principles density 
functional formalism or by constructing a semi-empirical Hamiltonian to compute the electronic 
eigenstates, total energy, and response functions to compare with experiments~\cite{Martin}. 
For amorphous materials, however, such an approach brings only a limited success. The determination of 
a ground state configuration (or an approximated one) of a glassy system involves 
identification of the global (or a low lying local) minimum in the multidimensional energy landscape, 
which is a very difficult problem in computational sciences. The complexity of 
the problem increases enormously with the increase of system size, and can 
be solved for a few simple potentials under strict mathematical conditions. 
The brute force applications of first-principles energy functional to model large amorphous systems 
is not only hopelessly difficult, but also uninspiring and time consuming.  Most of the {\it ab initio} methods 
asymptotically scale as $O(n^3)$ or even faster that seriously constrain the size of the simulation cell and 
total time to be studied. The so-called order-N counterpart of such 
approaches can improve the situation; however, for disordered materials the use of such methods are frankly 
very risky. The latter exploits the locality of quantum mechanics either via short-range nature of 
Wannier functions or truncation of density matrix of the system, but such requirements are rarely 
satisfied at the beginning of simulation for an arbitrary starting configuration. The prohibitive 
scaling constrains the system to evolve only for a limited time preventing it to explore the full 
solution space.  Except for small system size and for few good glass formers, direct {\it ab initio} calculations 
tend to produce unrealistic configurations of the material, and the situation is particular exacerbated 
for (amorphous) materials that do not display glassy characters (e.g. amorphous silicon).

In recent years there have been a number of methods that make use of the available knowledge 
of the materials (either experimental or otherwise) in the simulation methodology. These 
`knowledge-oriented' approaches have been developed under the general class of techniques 
based on information paradigm. Under this wing, one combines the power of first-principles 
force-field with the available information of the material under study. The general 
strategy is to construct a series of guesses of the solution that are consistent with experimental 
observation, and to improve the solution either deterministically or stochastically via an 
appropriate force-field.  Examples of methods that are currently in use include 
`Decorate-and-Relax'~\cite{Tafen-DR}, `Block-MD'~\cite{Tafen-BMD} and `Experimentally constrained 
molecular relaxation (ECMR)'~\cite{Biswas-ECMR}, where prior knowledge of the materials are used 
to accelerate search procedure for finding a better minimum in the simulation methodology. 
Imposition of suitably constructed prior (information) reduces the volume of the search space by 
hierarchically constructing more probable solution space, and the system is driven to find the 
best solution in this subspace.  Since the system spends most of its time in the favorable subspace 
of the desired solutions, it is computationally more  efficient and it is likely to find a better 
estimate of the global minimum of the problem. 
The essential idea is to `throwing out' the irrelevant part of the solution space by judicious choice 
of constraints and experimental data available for the material. 
While there exists no general scheme for guessing a series of candidate solutions for a
complex multinary system, one can nonetheless construct a set of disordered networks that shares some of the 
characteristic features of the material (here {\it a}-SiC) obtained from experiments. 
This is done by introducing various priors in the network, which identify and sample the more probable part of 
the solution space associated with that priors.

Following this strategy, the starting point of our calculation is to build a series of binary random 
networks having some degree of local tetrahedral order. Each of the binary  networks is then given 
different geometrical and topological character by changing the bond length, angles, coordination and
different ring geometry. From a macroscopic point of view, all of these structure have almost identical 
tetrahedral character (of varying degree), but topologically they are quite different. A number of 
samples are generated with varying degree of chemical and topological disorder so that during the course 
of simulation the system has access to entire solution subspace of approximated tetrahedral ordering.  
The particular choice of ordering is motivated by the available 
experimental data from \SIC (such as structure factor and bond angle distribution) suggesting presence 
of significant tetrahedral-like ordering 
in the material. It is to be noted that the topological character of this local ordering is very different 
from the randomized {\it c}-SiC structure~\cite{note1}, which is often used as a starting configuration without 
much care. While an unbiased, completely random configuration would be ideally the best choice to build 
the model during the course of simulation, such a choice requires exploring a vast amount of solution space 
at the expense of computation time without the guarantee of identifying the correct minimum from the potentially 
correct solution set. The method can be made feasible for identifying the global minimum in 
small systems (such as clusters, macromolecules and some glasses)~\cite{Wales}, but for large systems the number of 
solutions diverge rapidly with system size. In the present work, the binary random networks are 
generated via modified Wooten, Winer and Weaire (WWW)~\cite{WWW} algorithm as proposed by Barkema 
and Mousseau\cite{Moose1, Moose2}.  The structural and topological properties of these binary 
networks are further modified so that the resulting networks span the subspace of the entire solution 
space that are consistent with available experimental data. In particular, we focus here on a 1000-atom 
model with 50-50 concentration of Si and C having the same density as that of experimental atomic 
density of crystalline 3C-SiC.

\section{Molecular dynamics simulation procedure}
In this work we have used the {\em ab initio} program {\sc SIESTA}~\cite{Siesta1,Siesta2,Siesta3}, 
which is capable of determining the electronic structure and properties of molecules, surfaces and 
bulk materials based on density functional theory (DFT). The Perdew-Zunger formulation of the local 
density approximation (LDA) is employed~\cite{PZ} along with the norm-conserving Troullier-Martins 
pseudopotentials~\cite{TM} factorized in the Kleinman-Bylander form~\cite{KB} to remove core 
electrons. To describe the valence electrons, a set of atomic orbitals basis are used consisting of 
finite-range numerical pseudoatomic wave functions of the form proposed by Sankey and Niklewski~\cite{SN}. 
We employ a single-$\zeta$ (SZ) basis set on all atoms (one $s$ and three $p$ orbitals per atom) for the 
present calculation. Although some tight-binding calculations have indicated the need for an improved 
basis set beyond single-$\zeta$, in this work we restrict ourselves to a single-$\zeta$ basis to minimize our 
computational effort.  A real space mesh equivalent to a plane wave cut-off of 100 Ry is used for
the evaluation of the Hartree and exchange-correlation energies. Owing to the large system size and the 
cubic scaling of the computational effort with system size, we implement the approximation by 
Harris~\cite{Harris}, which is a non-self consistent version of DFT based on the linearization of 
the Kohn-Sham equations. The eigenvalue equations are solved by full diagonalization of the Hamiltonian 
matrix. 

\begin{figure} 
\begin{center} 
\includegraphics[width=0.50\linewidth]{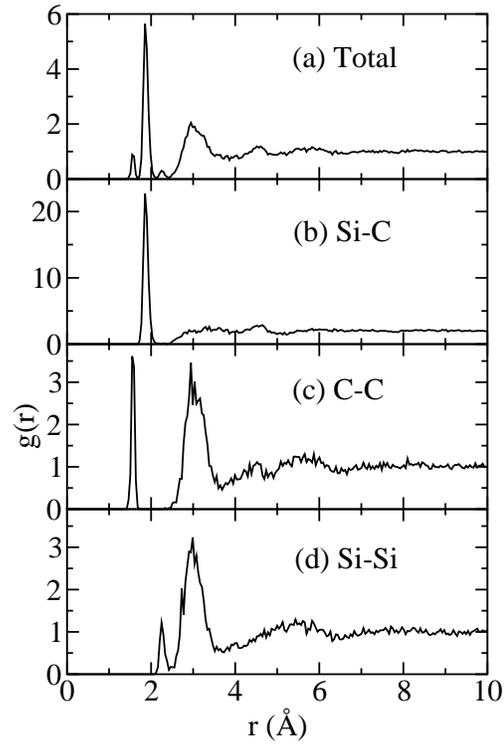}
\caption{
The total and partial radial distributions for the various pairs obtained from the 
model network. The distributions are all normalized for the purpose of comparison. 
} 
\label{rdf} 
\end{center} 
\end{figure}

\begin{figure} 
\begin{center} 
\includegraphics[width=0.50\linewidth]{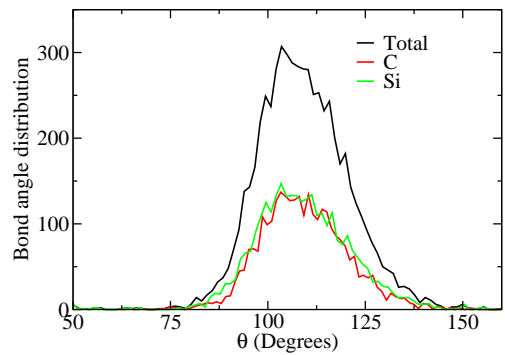}
\caption{(Color online) 
The bond angle distributions for \SIC network as discussed in the text. The distributions 
for angles centered at C and Si atoms are plotted along with the total bond angle 
distribution as indicated in the figure.} 
\label{bad} 
\end{center} 
\end{figure}

We performed a constant volume MD simulations using the Nos{\' e} thermostat with 
a time step $\Delta t$=1.0 fs. The simulation procedure consists of the following steps: 
(i) the temperature of the system is initially raised to 2500 K and is equilibrated for about 
0.5 ps (ii) the resulting system is then cooled down to 1000 K over a time period of 1 ps in 
steps.  (iii) We then equilibrate the system at 1000 K for 0.5 ps (iv) followed by cooling 
to 300 K over a time period of 1 ps. (v) The system is further equilibrated at 300 K for 0.5 ps,  (v) and 
finally it is relaxed at 0 K to the nearest minimum using the conjugate gradient (CG) method until the 
maximum atomic force on each of the atoms is less than 0.03 eV/{\AA}. These steps are then repeated for 
each of the candidate solutions mentioned in Section 3. 

\section{Results and discussions}

\subsection{Local structure and bonding environment}

We begin our discussion by focusing on the local structure and the bonding environment of the atoms 
via 2- and 3-body correlation functions. The correlation functions are calculated as radial 
and bond angle distributions providing information on the nature of short-range order 
in the network. For this 
binary system, we have studied three partial radial distributions of the components: C--C, 
Si--C and Si--Si. Together with the reduced 3-body correlation, the bond angle distribution, 
the local network properties can be studied and compared to the structural data from experiments. 
In figures~\ref{rdf}(a)-(d), we present the normalized total and partial 
radial distribution functions (RDF).  
The Si-C first maximum peak in fig.~\ref{rdf}(b) is clearly seen at 
around 1.9 {\AA}, which is in agreement with experiments~\cite{Kaloyeros, Ishimaru1} 
and previous theoretical calculations~\cite{Finnocchi, Kelires, Tersoff 94, Mura, 
Ivashchenko, Malerba, Yuan, Gao2, Rino}. Similarly, the C--C and Si--Si first 
maxima at 1.55 {\AA} and 2.25 {\AA} respectively, are also in agreement with 
experimental data~\cite{Ishimaru1}.  The total RDF in fig.~\ref{rdf}(a) clearly indicates a 
dominant Si-C peak implying presence of strong chemical order in the network. A comparison 
of the peak heights of the C-Si and C-C from the partial RDFs suggests that the network is 
dominated by chemical order as far as the RDFs are concerned. 

Further information about the local environment can be obtained via studying the 
coordination between Si and C atoms. For this purpose, we have used the location of the 
first minimum from the partial radial distributions as cut-off distances to define the 
first coordinations shells. In particular, the Si--C, C--C and Si--Si nearest 
neighbor cut-offs have been found to be $\rm R_{Si-C}$=2.25 {\AA}, $\rm R_{C-C}$=1.9 {\AA} 
and $\rm R_{Si-Si}$=2.5 {\AA}.  The bond angles centered  at Carbon and Silicon atoms are also 
calculated using the same cut-offs parameter.  In fig.\ref{bad}, we plot the bond angle 
distributions for the C- and Si-centered angles along with the total bond angles.  The average 
bond angle for the total distribution has been found to be $\langle \theta \rangle$ = 109$^\circ$ 
with a root mean square (RMS) deviation of about $\Delta\theta$=12.1$^\circ$. 
The corresponding values for the partials are $\langle \theta \rangle_{\rm C}$ = 
109.2$^\circ$, $\Delta\theta_{\rm C}$=11.3$^\circ$  and $\langle \theta \rangle_{\rm Si}$ = 108.7$^\circ$,  
$\Delta\theta_{\rm Si}$=12.9$^\circ$. This suggest that the final network continues to display 
tetrahedral character of its its crystalline counterpart. Together with the short range chemical 
order (manifested in the partial RDFs) and atomic coordination, the width of the bond angle 
distributions determine the overall quality of the network.  A large value of the width $\Delta\theta$ 
is an indicative of high degree of disorder that affects the electronic properties by 
introducing gaps states in the electronic spectrum.  Since electronic properties play an important 
role structure evaluation, a comparison with the other models in the literature would be relevant 
at this point.  Devanathan \etal~\cite{Gao2} obtained a value of $\Delta\theta_{\rm C}$=16$^\circ$ 
and $\Delta\theta_{\rm Si}$=36$^\circ$, whereas Rino ~\etal~\cite{Rino} reported a value of 
$\Delta\theta_{\rm C}$=15$^\circ$ and $\Delta\theta_{\rm Si}$=16$^\circ$. 

\begin{table}
\caption
{\label{tabone} 
Comparison of various structural parameters between this work 
and the previous experimental and theoretical works.  $\langle N_{C} \rangle$ and $\langle N_{Si} \rangle$ 
are the partial average coordination numbers of C and Si respectively, while $\langle N \rangle$ is 
the total average coordination number.  The chemical disorder parameter is denoted by 
$\chi=n_{CC}/n_{SiC}$, where $n_{\mathrm CC}$ and $n_{\mathrm SiC}$ 
are the percentage 
of C--C, Si--C 
bonds respectively. 
}  

\begin{indented}
\lineup
\item[]\begin{tabular}{llccc}
\br
&$\chi$&$\langle N_{C} \rangle$&$\langle N_{Si} \rangle$&$\langle N \rangle$\\                             
&       &                      &                        &                   \\                           
\br
Present work                         &0.083&3.98&4.01&4.00\\
{Kaloyeros 1988}$^{\rm a}$   &0.00 & & &3.99\\
{Finocchi 1992}$^{\rm b}$   &0.6&3.85&3.93& 3.89\\
{Kelires 1992}$^{\rm b}$     &0.5&3.46&4.02&3.74\\
{Tersoff 1994}$^{\rm b}$    &0.50  &4.00& & \\
{Mura 1998}$^{\rm b}$        &0.60  &    &    &  \\
{Ivashchenko 2002}$^{\rm b}$$^\ast$ &0.027&3.94&3.96&3.95\\
{Gao 2001}$^{\rm b}$         & & & & \\
{Rino 2004}$^{\rm b}$        &0.00& & &3.79\\
\br
\end{tabular}
\item[] $^{\rm a}$ Experiment
\item[] $^{\rm b}$ Theory
\item[] $^{\ast}$ Configuration labelled TB-216D in Ref.~\cite{Ivashchenko}.
\end{indented}
\end{table}

\begin{table}
\caption 
{\label{tabtwo} 
Irreducible ring statistics based on the nearest neigbbour cut-off defined in the text.}
\begin{indented}
\lineup
\item[]\begin{tabular}{cccccccc}
\br
Ring size& 3 & 4 & 5 & 6 & 7 & 8 & 9\\                             
Rings per atom& 0.002 & 0.015& 0.432& 0.749& 0.495& 0.146& 0.040\\                             
\br
\end{tabular}
\end{indented}
\end{table}

\begin{figure} 
\begin{center} 
\includegraphics[width=0.8\linewidth]{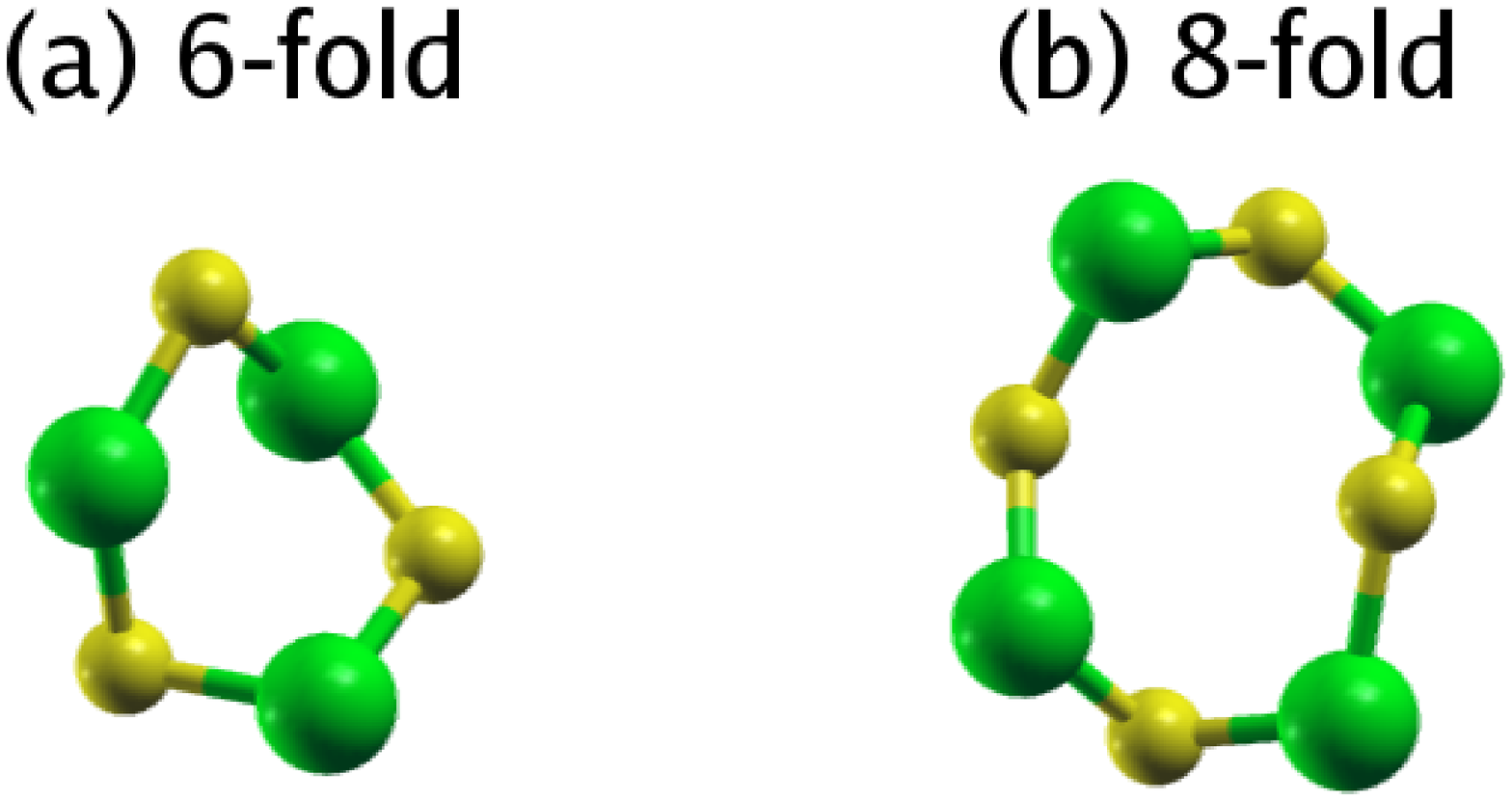}
\caption{(Color online) A representative 6- and 8-member irreducible rings present in the network. The Carbon and 
Silicon atoms are shown in yellow (small) and green (large) respectively. } 
\label{rings} 
\end{center} 
\end{figure}

A question of considerable importance and somewhat controversial is the extent of chemical disorder 
present in amorphous SiC network.  Unlike amorphous silicon where {\em sp$^3$} hybridization is the mechanism 
of bond formation between the neighboring atoms, the carbon chemistry permits C atoms to bond 
via {\em sp$^1$, sp$^2$} and {\em sp$^3$} hybridization making the local structure much more complex in 
disordered environment. Following Tersoff~\cite{Tersoff 94}, we define a short range order parameter
$\chi$  as the ratio of the number of C--C and C--Si bonds present in the network. Denoting $n_{\mathrm AB}$
(where A/B = Si, C) as the percentage of A--B bonding, we have found the values of C--C, Si--C, and 
Si--Si as 7.1\%, 85.5\%, and 7.4\%  respectively.  Similarly, the percentage $C_k$ of atoms that 
are $k$-fold coordinated have been found to be 2.6\% (3-fold), 95.3\% (4-fold), 2.0\% (5-fold), and 
0.1\% (6-fold) respectively.  This shows that the total coordination defect concentration is only 
about 4.7\%, implying a minor deviation from an ideal 4-fold coordination.  The percentage of 
C--C, Si--C and Si--Si bonds are respectively $n_{\mathrm CC}$=7.1\%, $n_{\mathrm SiC}$=85.5\% and $n_{\mathrm SiSi}$=7.4\%, 
giving $\chi$ of about $\chi$=0.083. A comparison of the various structural parameters with earlier 
studies is presented in Table~\ref{tabone}. 

In order to get the topological connectivity of the network, the type and the number of rings present are 
structure are computed.  The irreducible ring statistics for various $n$-member rings are computed and 
the results are reported in Table~\ref{tabtwo}. We define an irreducible ring as one such that it cannot 
be further partitioned into smaller rings. In other words, for a ring of size $n$, the minimum path starting from 
a given atom and to coming back to the same atom in an irreversible manner consists of $n$ hops. 
A 6-member ring, which is the only ring in crystalline SiC (with ZnS structure), is dominant in our network followed 
by 7- and 5-member rings. This again implies the presence of tetrahedral character as already observed via bond 
angle distribution.  Owing to few homonuclear bonds, almost all the rings consist of alternating Si and C atoms. 
In fig.~\ref{rings}, we have shown two representative 6- and 8-member rings. 

We summarize the structural properties by noting that 1) the network has small RMS width of the bond angle 
deviations compared to models in the literature, and 2) having low coordination defect. Together with these, 
the ring statistics further confirms that the network is not strained and, therefore, should have good electronic 
properties. The latter aspect of the model is addressed in the next section in this paper. Furthermore, we have 
also noted, contrary to previous calculations~\cite{Finnocchi,Kelires}, no graphene-like structure (3-fold coordinated planer 
carbon region) with $\theta$=120$^\circ$). This is not surprising in view of the fact that our model has a very 
few C-C wrong bonds.

\begin{figure} 
\begin{center} 
\includegraphics[width=0.60\linewidth]{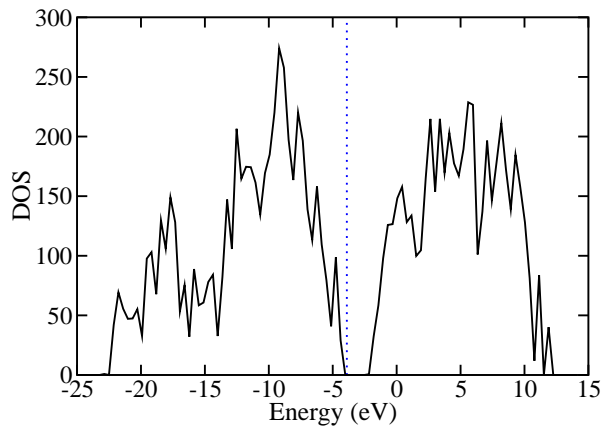}
\caption{ The electronic density of states for \SIC network obtained from a first-principles {\sc Siesta} Hamiltonian 
discussed in the text. The eigenvalues are broadened using a gaussian width of 0.01 eV. The Fermi level is 
indicated as a dashed vertical line in the figure. 
}
\label{edos} 
\end{center} 
\end{figure}

\subsection{Electronic Properties}
While a great deal of information can be obtained from studying structural properties of the network 
via 2- and 3-body correlation functions and local coordination analysis, the electronic spectrum 
provides a more refined aspect of the local environment. In particular, the degree of local disorder is reflected in 
the local electronic density of states (l-EDOS) via formation of gap states and in the nature of the 
band tails of the spectrum.  A high degree of local disorder causes the gap narrow and noisy, which is 
difficult to gauge from only partial RDFs and bond angle distribution. Partial coordination number 
analysis does provide some information on the nature of eletronic density of states (e.g. presence of 
dangling bonds), but for a detailed understanding of eletronic properies and the nature of the wave 
functions near the band gap(s), one needs to calculate eletronic density of states of the model network. 
This is particularly important for models obatined via empirical potentials, but care must be taken 
into account to make sure the electronic density of states is reasonably good, and reflects the correct local 
chemistry.  The electronic density states for our model is obtained via diagonalization of the density 
functional Hamiltonian from {\sc Siesta}. 

In fig.~\ref{edos}, we have presented the gaussian-broadened electronic density of states (EDOS) 
for the single particle Kohn-Sham energy eigenvalues. Several important observations are in order. 
First, the model shows the presence of a clean spectral gap. For a finite system, the gap size can be 
estimated as the difference between the highest occupied molecular orbital (HOMO) 
and the lowest unoccupied molecular orbital (LUMO).  The exact value of HOMO-LUMO gap for our model is found to be 
1 eV, which is about 1.2 eV smaller than the experimental bandgap for 3C-SiC of 2.2 eV~\cite{gap_exp}. 
However, as observed from the fig.~\ref{edos}, the average gap size is greater than 1 eV and is 
much closer to the experimental value taking into account some defect states that are close towards the 
main conduction band. It is noteworthy that even though LDA underestimates the size of the gap, our model 
does produce the presence of almost clean gap in the spectrum. This is a further indication that our method correctly identifies 
low lying minima in the configuration space. We want to emphasize that this is an important aspect of our 
method despite being approximate in nature. 

\begin{figure} 
\begin{center} 
\includegraphics[width=0.60\linewidth]{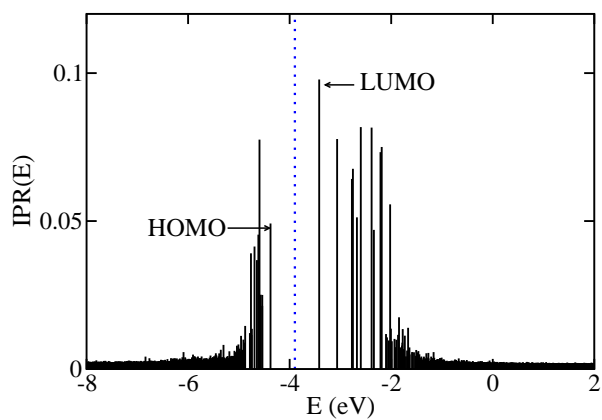}
\caption{ 
Inverse participation ratio (IPR) for the 1000-atom {\it a}-SiC model as a function of 
electronic eigenvalues. The highest occupied (HUMO) and the lowest unoccupied molecular 
orbitals are shown in the figure. The large IPR values reflect the localized nature of the 
states in real space. }
\label{eipr} 
\end{center} 
\end{figure}

\begin{figure} 
\begin{center} 
\includegraphics[width=0.35\linewidth]{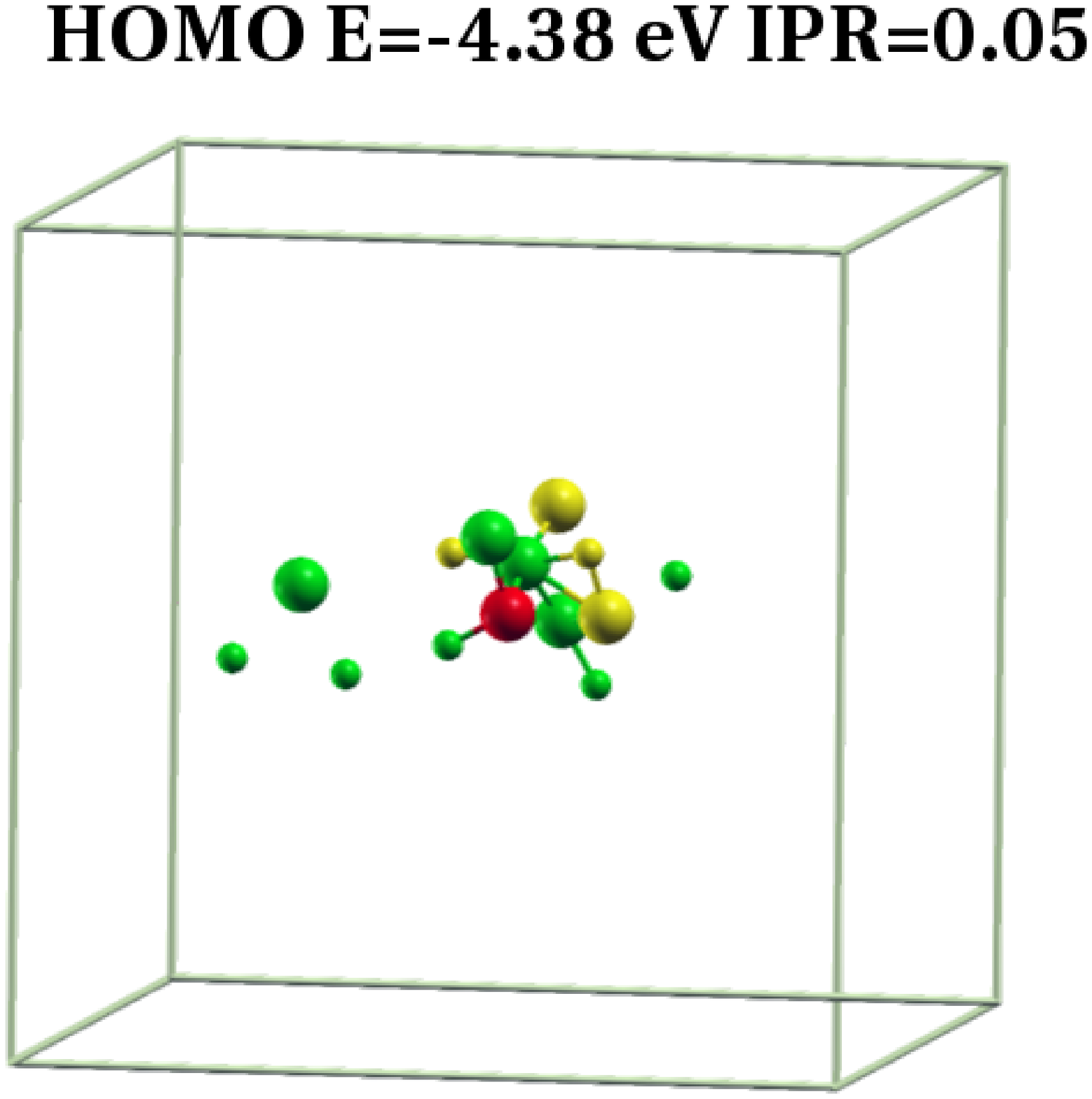}
\includegraphics[width=0.35\linewidth]{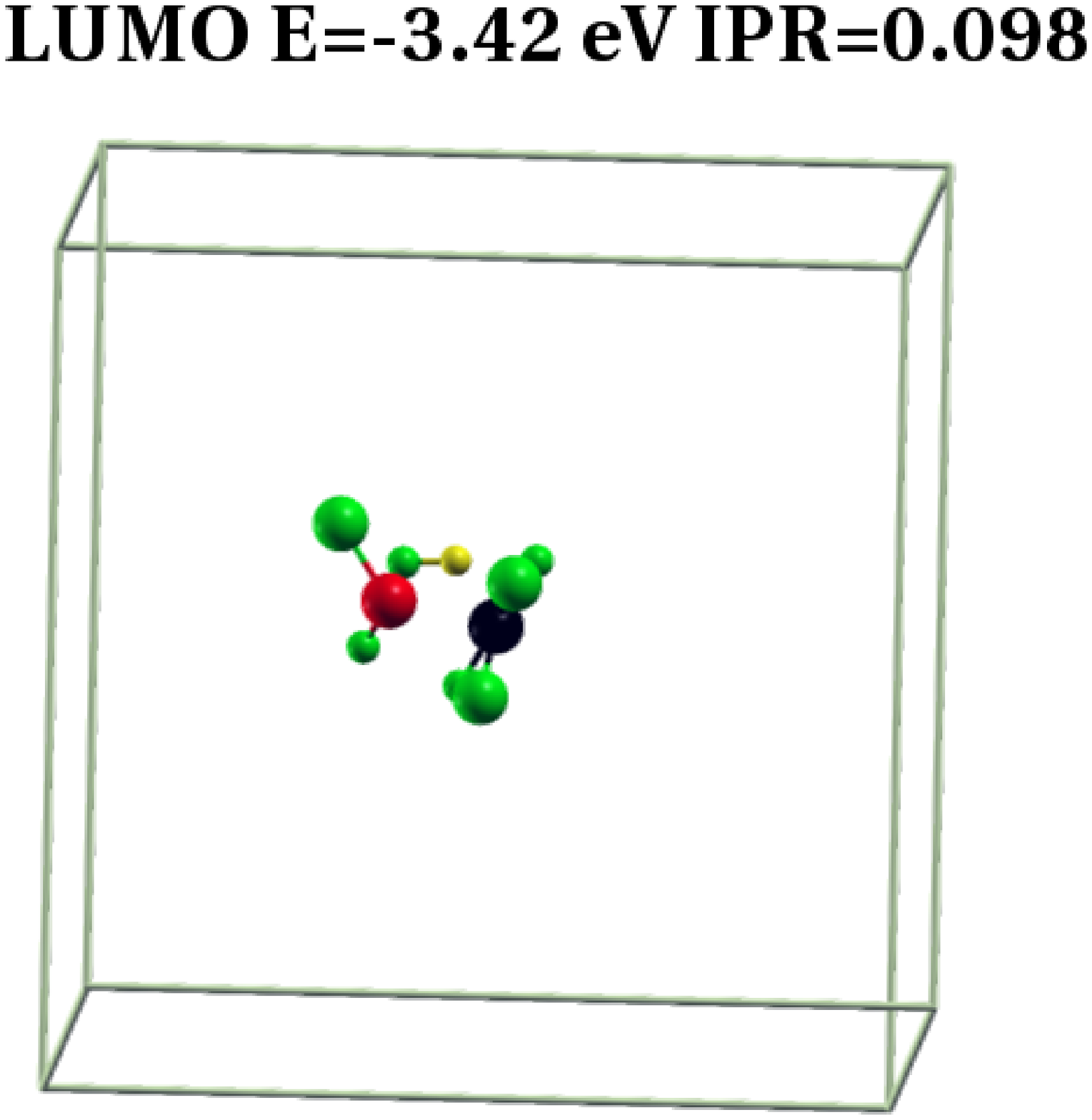}
\caption{(Color online) 
Spatial character of the localized HOMO (left) and LUMO (right) eigenstates.  Silicon atoms are shown somewhat 
larger than the C atoms. The following color code is used to map the Mulliken charge $q$ 
for the states at each atomic site: black ($q>0.20$), red ($0.10<q\le0.20$), gold ($0.05<q\le0.10$) and 
green ($0.01<q\le0.05$). 
}
\label{ipr_homo_lumo} 
\end{center} 
\end{figure}

A further characterization of our model comes from the nature of the band tail states. In an amorphous 
system, one expects the band tail states to show some signature of localization due to topological and 
chemical disorder in the network. This is independent of the presence of defect in the system.  To study this, one often uses 
an {\it ad hoc} measure for eigenstate localization known as the {\it inverse participation ratio} 
(IPR). In a simplest way, IPR approximately measures the {\it reciprocal} of the number of atoms 
that participate in localizing an eigenstate. While IPR is not the best way to quantify degree 
of localization, it does provide some qualitative idea about the spread of an eigenstate 
over the atoms. For a multi-band Hamiltonian, given a normalized eigenstate $\psi$ with eigenenergy $E$, 
we first compute the 
Mulliken charge contribution from each atom $q_i$ by projecting the eigenvector onto the orbitals centred on 
atom $i$ and then sum over the orbitals to get the total contribution. The IPR can be expressed as, 

\be
{\rm IPR}(E)= \sum_{i=1}^{N}q_{i}^{2}
\label{eqn2}
\ee

where $N$ is the total number of atoms. 
If the eigenstate $\psi$ is completely localized, then only a single 
atom participates in localizing $\psi$, yielding IPR($E$)=1.  Otherwise, if 
$\psi$ is completely delocalized, then all atoms participate in localizing $\psi$, giving 
IPR($E$)=1/$N$.  Thus large values of IPR generally correspond to localized states and vice versa. 
In fig.~\ref{eipr}, we plot the electronic IPR versus energy eigenvalue in the vicinity of the 
Fermi level. Each spike is located at an energy eigenvalue. The values of the IPR clearly show 
that the states near the band gap are more localized than of the states away from the gap. 
A pictorial representation often helps better to understand this behavior and is presented in 
fig.~\ref{ipr_homo_lumo}. To this end, we assign different colors to each site according to its 
Mulliken charge contribution for a given eigenstate.  We then depict the spatial feature by 
showing a fraction (70\%) of the total charge for the HOMO and LUMO states in the plot. 

It is clear from the fig.~\ref{ipr_homo_lumo} that the eigenfunctions (corresponding to HUMO and LUMO) 
mostly confined to a small cluster of atoms in space.  While the size of the cluster is, to an extent, 
dependent on the chosen fraction of Mulliken charge, it's nonetheless confirms that the localized nature 
of the states in real space. 
For the occupied state, a silicon atom (large and colored red in the plot) has the largest contribution of about 
14\% of the unit charge.  The atom is 4-fold coordinated, and have 2 homonuclear and 2 heteronuclear 
bonds. For the LUMO state, the largest contribution to the Mulliken charge is about 25\% from a silicon 
atom (colored black). The atom is also 4-fold coordinated with 2 homonuclear and 2 heteronuclear bonds.

\subsection{Vibrational Spectrum}   
The vibrational density of states (v-DOS) of the model can been studied by computing the vibrational eigenmodes 
and eigenfrequencies. Since the energy associated with a typical vibrational degree of freedom is much lower than 
that of the eletronic, a slight deviation in the local geometry can cause a significant deviation in the 
vibrational spectrum and the modes of the vibration of the atoms. Such changes often can be difficult to observe 
in electronic density of states, and therefore, for a full characterization of the model one need to ascertain the 
vibrational eigenmodes and the nature of vibrations. In this work, we address this issue by computing the 
modes via direct diagonalization of the dynamical matrix. Once the eigenmodes are available, the corresponding 
vibrational IPR can be calculated for each of the eigenfrequencies and the nature of the modes can be studied. 

The dynamical matrix elements are constructed by successively displacing 
each atom in the fully optimized supercell along the positive and 
negative direction of the three orthogonal axes, and computing the 
resulting atomic forces within the harmonic approximation. 
Unlike the electronic IPR, we define here the vibrational IPR in a somewhat different way. 
Denoting a normalized vibrational 
eigenmode with frequency $\omega$ by ${\bf \Phi}=({\bf \Phi}^{1}, {\bf \Phi}^{2},\ldots,{\bf \Phi}^{N})$, where 
${\bf \Phi}^{i} =(\Phi^{i}_{x}, \Phi^{i}_{y}, \Phi^{i}_{z})$ and $i$ is the atomic index, the IPR is 
defined as 
\be
{\rm IPR}(\omega)= \sum_{i=1}^{N}({\bf \Phi}^{i}\cdot{\bf \Phi}^{i})^{2}
\ee
For an ideally localized ${\bf \Phi}$, only one atom contributes to the vibrational amplitude 
and so IPR($\omega$)=1. Similarly, for an ideally delocalized ${\bf \Phi}$, all atoms contribute 
uniformly to the amplitude yielding IPR($\omega$)=1/$N$.

\begin{figure} 
\begin{center} 
\includegraphics[width=0.60\linewidth]{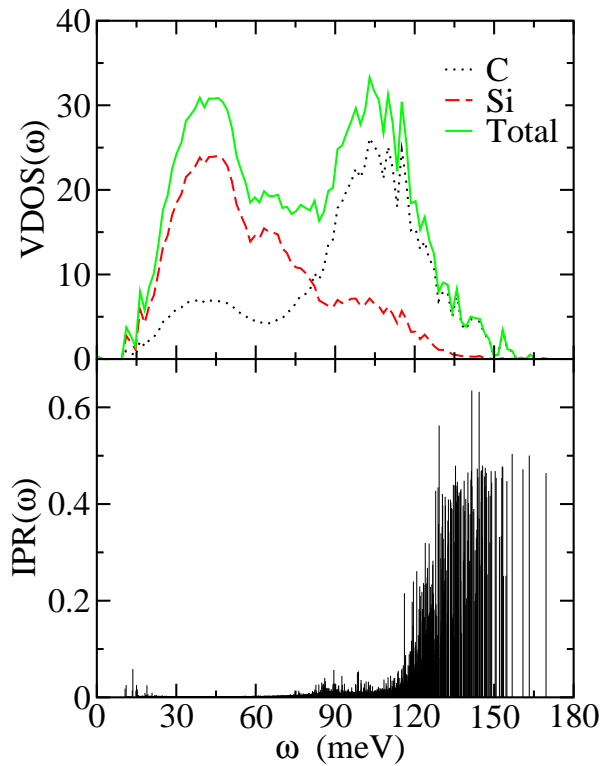}
\caption{ (Color online upper panel)
Atom-projected and total vibrational density of states (upper panel) and vibrational 
IPR (lower panel) for the 1000-atom \SIC model.  The majority of the high frequency 
vibrational modes are localized around C atoms of C--C and Si--C bonds. 
}
\label{vipr} 
\end{center} 
\end{figure}

In fig.~\ref{vipr} we show, in the upper panel, the plots for the atom-projected (weighted 
by the atomic contribution to the amplitude of the eigenstate conjugate to 
a given eigenfrequency) and total v-DOS. The corresponding vibrational IPR is 
plotted in the lower panel. 
The v-DOS clearly shows that the low frequency bands are dominated by 
Si while the high frequency bands are dominated by C. The overall shape 
and peak positions of the partial and the total v-DOS are in agreement with recent calculations by 
Vashishta~\etal~\cite{Priya} using an empirical potential. The IPR for the vibrational modes
clearly distinguish two different type of modes: low frequency extended modes primarily arising from 
the Si atoms and the high frequency localized modes originate mostly from the C atoms. 
The nature of the vibrational mode can be obtained by studying the corresponding eigenmode and the 
contribution from each of the atoms. The most localized vibrational mode appears at frequency 142 
meV having IPR value 0.634. A real space analysis reveals that the vibration corresponds to this 
mode is a Si--C stretching mode with 78\% of the mode confined on the C atom, and about 12\% on the 
Si atom leaving behind the rest 10\% on the neighboring atoms. Similar analysis of the other localized 
states suggest all the high frequency states are localized and are centered on C atoms.

\section{Conclusion} 

We present an approximate first-principles molecular dynamics simulation of \SIC within the 
density functional formalism coupled with a priori information obtained from experimental 
data. 
The method consists of generating a set of smart structural configurations as candidate 
solutions starting from a generic binary tethedral network and importance sampling the 
configuration subspace by incorporating characteristic structural, topological, and 
experimental information. 
Each of the samples is then used in our {\it ab initio} constant temperature molecular 
dynamics to find the minimum energy configuration, and the global minimum of this set used 
as a final solution in our work. We have studied structural, electronic and 
vibrational properties for 50-50 \SIC by using a localized basis first-principles molecular dynamics 
method. The degree of chemical disorder, expressed as a ratio of the C--C to the Si--C 
bonds is found to be approximately 0.083. The percentage of C--C, Si--C and Si--Si bonds 
in the network are observed to be 7.1\%, 85.5\% and 7.4\% respectively. The coordination 
defect concentration found in our model is low and is about 4.7\%, while the average bond angle and 
its RMS width are about 109$^\circ$ and 12.1$^\circ$ respectively. The electronic spectrum 
reveals a reasonably clean gap (HOMO-LUMO) of size about 1 eV taking into account the 
defects states in the gap. Inverse participation ratio (IPR) analysis indicates that 
the electronic band tail states are fairly localized with almost central atom being 4-fold 
coordinated, and having at least one wrong bond. Similar analysis of the vibrational spectrum within 
the harmonic approximation suggests that the low frequency bands are dominated 
by Si vibrations whereas the C atoms mostly involve with the high frequency vibrational 
modes. A real space analysis of the high frequency vibrational modes (identified via IPR) shows that 
the vibrations are mostly stretching in character with larger contribution coming 
from the C atoms. 

\ack
The authors thank Normand Mousseau (Univ. of Montreal) and Gerard Barkema 
(ITP, Utrecht) for providing binary continuous random networks that are 
used for smart structure generation. PB acknowledges the support of the 
University of Southern Mississippi under Grant No. DE00945.

\end{document}